\documentclass[11pt]{article}

\usepackage[utf8]{inputenc}
\usepackage{graphicx,booktabs,amsmath,amssymb,url,xcolor,authblk,natbib}
\usepackage[margin=1in]{geometry}
\usepackage[colorlinks=true,allcolors=blue]{hyperref}

\newcommand{\rqA}{RQ1}   
\newcommand{\rqB}{RQ2}   
\newcommand{\nscicat}{18{,}247} 

\title{\textbf{The Reciprocal Impact of Science and Software:\\
A Cross-Corpus Analysis of How Research Shapes Software and\\
Software Enables Research}}

\author[1]{Audris Mockus}
\affil[1]{University of Tennessee, Knoxville \quad \texttt{audris@utk.edu}}
\date{Draft \today}

\begin{document}
\maketitle

\begin{abstract}
Software and scientific knowledge co-evolve, yet they are catalogued in
separate corpora that rarely speak to one another: software lives in version-control
ecosystems, science in bibliographic databases. We bridge them at global scale by
linking \emph{World of Code} (the near-complete mirror of public version-control
history) to \emph{Semantic Scholar} and \emph{OpenAlex} through a typed cross-corpus
graph of 69.8M edges spanning eight relation types (paper$\to$software mentions,
software$\to$paper citations, software$\to$software dependencies, authorship,
affiliation, and identity bridges). Anchoring on a curated set of 18{,}247 \emph{science}
repositories, we ask two reciprocal questions: (\rqA) what is the impact of \emph{science
on software} (which papers and which scholars most shape research software?), and
(\rqB) what is the impact of \emph{software on science} (which research software most
enables published work?). To test whether this Science-Software Supply Chain (S3C)
conceptualization is
feasible and what it can reveal, we conduct several basic investigations rather than claim a
definitive measurement. Across them, the two directions appear to illuminate \emph{different
and complementary} strata of the ecosystem: the literature's reach into software is dominated
by a reproducibility/packaging layer (nf-core, Nextflow, Singularity, Bioconda) and
heavily used sequence-analysis tools, whereas software's reach back into science is
\emph{proxied} by a largely invisible machine-learning / data-science \emph{infrastructure}
tier (PyTorch, seaborn, NLTK) on which the broader software ecosystem depends. We are explicit
that this second direction is carried by a software$\to$software reuse signal (ecosystem
dependency in-degree, counted over \emph{all} World-of-Code projects rather than the science
seed alone) because the direct paper-names-software channel proves too sparse to rank: an
independent human-curated gold benchmark links none of its 65 in-scope cases
(\S\ref{sec:validation}). Dependency reuse therefore stands in as a proxy for scientific
enabling rather than measuring it directly. In our data broad ecosystem adoption
is at most weakly coupled to scholarly citation count and to
popularity (stars, Spearman $\rho=0.36$): the near-top-ranked package may have two orders of
magnitude fewer stars than a less-reused one.
We read these as feasibility evidence that the two disjoint ledgers can be brought into a single,
more transparent view, giving the science-of-science literature a software-aware,
dependency-grounded signal that citation counts alone cannot recover, and suggesting such
transparency is both attainable and needed.

Our most cautionary finding is about the measurement itself. The reuse--citation coupling
flips sign and confidence across two reasonable ways of pairing a repository with a citation
count: through the papers that \emph{name} it ($n=137$, $\rho=0.05$, CI straddling zero, read
as decoupled) versus through the DOIs a repository \emph{declares} for itself ($n=1{,}067$,
$\rho=0.13$, CI $[0.07,0.19]$, a weak positive coupling). Neither panel is wrong; the gap
between them is the point. With science--software linkage this sparse and selection-biased,
the sign of a headline correlation is an artifact of which gap one tolerates, so we report
both and refrain from a strong decoupling claim.
\end{abstract}

\section{Introduction}
\label{sec:intro}

\paragraph{New instruments, new science.}
Scientific breakthroughs have always followed better instruments: science flows from ideas
and from the means to \emph{measure}, not from data as such. The telescope and the microscope
opened whole disciplines not by accumulating observations but by making previously invisible
phenomena measurable. The instrument has since changed form. It is no longer only a physical
apparatus but an \emph{operational} one: large-scale data collection and analysis enabled by
the big-data turn, the software infrastructure that transforms the massive output of real
instruments into usable signal, and now the capacity of AI to process data and code at scale.
This same progression has produced a new class of instrument trained on the scientific
enterprise itself: corpora such as World of Code, Semantic Scholar, and OpenAlex that
render the entire software-and-literature record observable. They are what let us treat the
Science-Software Supply Chain (S3C) below as an \emph{organizing lens} (a way to ask which artifacts feed
which) rather than only a figure of speech, even though a full bill-of-materials traversal of
that chain remains future work.

\medskip
Modern science runs on software, and a great deal of software is produced by and for
science. But the two are recorded in disjoint ledgers. Bibliometrics counts papers and
their citations; software metrics count stars, forks, and downloads. Where the two meet
(a paper that names a tool, a repository that cites the method it implements, a library
that silently underpins a thousand analyses) the signal is fragmented across corpora and
mostly lost to formal measurement~\citep{howison2016softwarelit,istrate2022czidataset}.

\paragraph{The Science-Software Supply Chain.}
Our overarching goal is to treat scientific production as a \emph{supply chain} in which
papers and software are co-equal goods. The software-engineering community has learned to
view code as a \emph{software supply chain}~\citep{fse19,isectut22,
nasac17}: a directed network of dependencies, builds, and
contributors through which value (and risk) flows from upstream components to downstream
products, and whose centrality structure determines how much each contributor's output
matters~\citep{centrality23}. Science is the same kind of network, only larger: a
finding flows from data and methods into software, into the papers that use that software,
into the next generation of tools that operationalize those papers, and onward. Yet today
this chain is recorded in two disjoint ledgers, so its provenance is unobservable. As in any
supply chain, \emph{visibility and transparency are prerequisites for managing risk and
improving efficiency}~\citep{woc19url}: an analyst who cannot see that thousands of
analyses rest on a single unmaintained ``hidden hero'' library cannot price that fragility:
reproducibility risk, single-maintainer bus factor, dependency
abandonment~\citep{abandon25}, or orphaned vulnerabilities propagated through
copy-based reuse~\citep{vuln22}; a funder who cannot see which papers a tool enabled,
or which tools a finding seeded, cannot allocate credit or investment efficiently. A software bill-of-materials (SBOM)
makes a product's components auditable; the S3C would need the analogous
artifact: a global, identity-resolved \emph{bill of materials for scientific production}
that links every paper to the software it uses and produces, and every piece of software to
the literature it grounds and the developers who build it. Whether such an artifact can be
\emph{built and measured} at all, and what it reveals once it exists, is the question this
paper investigates, as a first step toward the transparency the chain requires, not as a
finished accounting of it.

This paper takes first steps toward measuring that meeting point in both directions, at the
scale of essentially all public open-source software. Three developments make such a study possible only now.
First, World of Code has matured into a near-complete, continuously updated mirror of
public version-control history with a resolved author-identity layer~\citep{woc19,AZBZM19},
so software and its developers can be enumerated globally rather than sampled from a single
forge or ecosystem. Second, the scholarly record has opened: Semantic Scholar and OpenAlex
expose the citation graph, author identifiers, and full-text--derived mentions
programmatically, turning paper$\leftrightarrow$software linkage from a manual annotation
task into a join. Third, the science-of-science
agenda~\citep{fortunato2018sciofsci,wang2021sciofsci} has established the questions (how
ideas, artifacts, and people accrue and transmit impact) but has pursued them almost
entirely on publications, leaving software as an unmeasured term. The conjunction of a
global software census, an open scholarly graph, and a maturing measurement agenda makes a
reciprocal, software-aware impact study feasible for the first time.

\paragraph{Research questions.}
\begin{itemize}
  \item[\textbf{\rqA}] \emph{Impact of science on software.} Which scientific papers and
        which scholars propagate most broadly into research software? How grounded in the
        peer-reviewed literature is research software?
  \item[\textbf{\rqB}] \emph{Impact of software on science.} Which research software most
        enables science, as measured by mentions in the literature and by being
        depended upon across the software ecosystem? Does this ``hidden infrastructure''
        coincide with what citation counts or popularity would surface?
\end{itemize}

\paragraph{Contributions.}
(i) A reframing of scientific production as a \emph{Science-Software Supply Chain} in which papers
and software are co-equal goods, and a typed cross-corpus graph (a first step toward a global bill of materials
for that chain) linking the near-complete open-source software universe to two major
bibliographic corpora, with identity bridges (author, DOI, repo-URL) reconciled to a
canonical space (\S\ref{sec:data}).
(ii) A pair of reciprocal, graph-path impact measures operationalizing \rqA{} and \rqB{}
(\S\ref{sec:methods}).
(iii) Feasibility evidence (\S\ref{sec:results}) that science$\to$software and
software$\to$science appear to surface distinct ecosystem strata, together with a cautionary
result that the strength of the reuse--citation coupling depends on how repositories are paired
with citations: it ranges from undetectable (papers that name a repository, $n=137$) to a weak
positive association (DOIs a repository declares, $n=1{,}067$), so we treat the coupling as
gap-sensitive rather than settled.
(iv) A time-aware treatment of AI coding-agent adoption: rather than regress decade-long
impact stocks on a 2023+ behavior, we identify what \emph{facilitates} adoption as a held-out
predictive task, and quantify why the post-adoption outcome study must wait
(\S\ref{sec:aiagents}).
(v) Implications for how the science-of-science community and research-software funders
should value software, grounded in a held-out validation against the human-annotated Softcite
gold corpus that bounds our linkage recall and quantifies a hard proprietary-tool ceiling
(\S\ref{sec:validation}, \S\ref{sec:discussion}).

\paragraph{In brief.}
Across three lenses, broad ecosystem reuse, scholarly citation, and popularity surface largely
distinct strata; stars do not substitute for dependency reuse.
Reading one global graph in three directions yields three almost-disjoint pictures of impact:
the literature reaches into software mainly through a reproducibility/packaging layer
(nf-core, Nextflow, Singularity), software reaches back into science mainly through an
invisible ML/data-science infrastructure tier (PyTorch, seaborn), and the scholars who
underpin the ecosystem are its toolmakers. None of these is recoverable from citation counts
or GitHub stars: dependency-based reuse is only weakly correlated with stars
(Spearman~$\rho=0.36$), and its coupling to scholarly citation is weak at best and unstable
under how the pairing is drawn (undetectable through naming links, $\rho=0.05$; weakly positive
through repository-declared DOIs, $\rho=0.13$), while the top repositories under each lens
overlap only slightly above chance. Finally, the same graph lets us follow a \emph{new} mode of software
production (AI coding agents) into the S3C. Because agent adoption is a
2023+ behavior while reuse and citations are decade-long stocks, we study its identifiable side:
\emph{what is associated with adoption}. Adoption is only weakly forecastable from pre-AI antecedents
(held-out AUC~$0.60$); the factors independently associated with higher odds are development intensity
and relative youth, not field or popularity. AI uptake is, so far, a property of active,
fast-moving projects rather than of any particular science.

\section{Related Work and Motivation}
\label{sec:related}

Our study sits at the intersection of four literatures: (i) the visibility and
citation of software in the scholarly record, (ii) large-scale extraction of
software \emph{mentions} from text, (iii) dependency- and usage-based measures of
software importance, and (iv) the science-of-science measurement agenda. Each has
made decisive progress, yet each stops short of the same thing: a global,
identity-resolved link between scientific papers and the actual software
artifacts they produce and consume, traversable in \emph{both} directions. We
review each thread, quantify what it establishes, and isolate the gap our
cross-corpus graph fills.

\subsection{Software is invisible and undercredited in the literature}
\label{sec:rel-invisible}
A foundational body of work documents that scientific software, though central to
research, is largely invisible in the published record. Howison and
Herbsleb~\citep{howison2011incentives}, tracing by hand the software-dependency
chains behind three focal papers in physics, structural biology, and
microbiology across 28 interviews, show that the scholarly reputation economy
rewards topical publications but not software use or maintenance: essential
infrastructure (the \texttt{ROOT4STAR} analysis library, data-production and
simulation code, and the SBGrid distribution serving $140+$ labs) is routinely
used yet never cited, ``invisible infrastructure work,'' and the resulting
incentive misalignment predicts over-production of independent packages and
under-production of collaborative ones. Howison and
Bullard~\citep{howison2016softwarelit} quantify this invisibility in a stratified
sample of 90 biology articles: only 65\% even mention software, and of 286
mentions just 31--43\% are formal citations (39\% cite a publication, 19\%
instrument-style, 43\% wholly informal); only 86\% of mentioned software is even
findable, version information is present in only 28\% of cases and a \emph{specific}
version is recoverable in just 5\%, source is exposed for 32\% and modification
permitted for 20\%, and a mere 18\% of packages request a citation at all. They
conclude that measuring software impact ``solely by searching for specific
citations has serious validity concerns,'' and explicitly call for the
infrastructure they lack: linking directly to the source-code repositories
developers use and recovering contribution through ``post-hoc examination of
commits and their authorship.''

Even \emph{formal} software citation, where it occurs, is not aggregatable.
Schindler et al.~\citep{multilevel2023dataquality} trace formal citations end to
end and find only 22.3\% are direct citations (69.6\% are proxy ``software
article'' citations that cannot identify a code base), only $\sim$66\% of direct
citations even identify the code base, and, for any cross-corpus
study, the bibliographic substrate is structurally blind to software: Semantic
Scholar silently drops 22.9\% of direct software references and injects wrong
metadata into $\sim$16\% of those it retains. Druskat et
al.~\citep{dontmentionit2024} corroborate that citation practice has not improved
since 2015 (name-only mentions now exceed 50\% of all mentions). Schindler et
al.~\citep{schindler2022rolesoftware} reinforce this from the corpus side: building a
knowledge graph of software mentions across PubMed Central (and, in
SoMeSci~\citep{schindler2021somesci}, a gold-standard mention knowledge graph for training
such extractors), they show software is pervasive
in the methods record yet recorded almost entirely as informal name-drops rather than
resolvable, citable artifacts, and that the form a mention takes varies sharply across
disciplines~\citep{pan2016disciplinary}, so any single citation- or mention-based proxy
samples software use unevenly between fields. Tools such as
CiteAs~\citep{citeas} attempt to recover the \emph{requested} citation for a piece of
software, but depend on the very metadata that is missing for most artifacts; the FORCE11
Software Citation Principles~\citep{smith2016softwarecitation} codify how software \emph{ought}
to be cited, but adoption remains the exception rather than the rule. The consistent
message is that the citation channel undercounts software severely and unevenly,
so impact measured from citations alone is unsound.

\subsection{Mining software mentions: detection solved, linkage not}
\label{sec:rel-mentions}
A second thread shows it is now feasible to extract software \emph{mentions} from
literature at massive scale. Istrate et al.'s CZ Software
Mentions~\citep{istrate2022czidataset} apply a SciBERT NER model (F1~0.92 on
SoftCite) to $\sim$20.7M biomedical papers, yielding $\sim$19.3M mentions and
1.12M unique strings across the PMC-OA corpus, which an
unsupervised pipeline collapses into 97{,}600 distinct software entities covering
78\% of software--paper links. The decisive limitation is that this literature
stops at the \emph{textual} mention and never reliably reaches the
\emph{artifact}. CZI links only a minority of mentions to a repository (roughly one
in six), by bare exact-name matching against uncurated GitHub, and its own evaluation flags 39 of 40
ambiguous links as GitHub name collisions. Druskat et
al.~\citep{dontmentionit2024} quantify the downstream cost on manual samples:
23--46\% of sampled mentions are not even software; of the automatic
repository links present, 65.4\% point to the \emph{wrong} software and only 20.5\%
of linkable mentions are linked at all, leading them to conclude that current
mention datasets cannot support repository mining (source, license, version
history, issue tracker, development process). Detection of the mention is solved;
the connection from mention to the real software object (its dependency network
and its developers) is not.

\subsection{Dependency and usage as an impact axis, decoupled from citation and popularity}
\label{sec:rel-dependency}
A third thread establishes that the \emph{importance} of software is decoupled
from the signals usually used to measure it. Druskat~\citep{software2019citationgraphs}
provides the formal frame: a citation graph that omits software and its
dependencies renders software's contribution structurally invisible, and the
lower-stack ``hidden'' dependencies (never published, carrying no citation
metadata) can be recovered only by mining software artifacts, not publication
records; but his account is a model with no instantiation. Li, Chen and
Yan~\citep{li2018lme4impact} show empirically why citation alone fails even for a
single heavily used package: \texttt{lme4}'s impact splinters across $10+$ citable
objects and $100+$ inconsistent cited forms, a single proxy captures only
$\sim$half the citations, and a dedicated software-citation index recovered only
$\sim$6\% of one object's true count. Brown et al.~\citep{biomed2024hiddenheroes}
operationalize the dependency axis as the corrective: over a paper-mention +
dependency network of biomedical software, Katz centrality and paper visibility
are nearly orthogonal (mention Gini $\sim$0.84--0.89), and concrete ``hidden
hero'' packages (\texttt{isoband}, 0 mentions; \texttt{vctrs}/\texttt{withr},
2 mentions; \texttt{tifffile}; \texttt{setuptools}) sit in the top $\sim$1\% of
centrality, while ``popular'' packages (GSVA, MAST: 2{,}000--3{,}000 mentions)
fall below the median in centrality. Dependency in-degree thus measures a kind of
importance that neither citations nor mention-counts (the scholarly analogue of
GitHub stars) capture. Yet Brown et al. confine the analysis to $\sim$17\% of
open-source packages across three managed ecosystems (PyPI/CRAN/Bioconductor) in
one domain, using \emph{declared} rather than \emph{observed} dependencies, and
measure impact only \emph{into} software; the reciprocal, global, cross-ecosystem
direction is unmeasured. Broad dependency-ecosystem studies~\citep{decan2019ecosystems} and
cross-ecosystem package indices such as Libraries.io and
Ecosyste.ms~\citep{nesbitt2024ecosystems} likewise map \emph{declared} package dependencies at
global scale, but in isolation from the scholarly record and without resolved developer
identity.

\subsection{The science-of-science agenda omits software}
\label{sec:rel-sciofsci}
Finally, the science-of-science field frames the data-driven study of scientific
production as the analysis of a single multiscale network of scholars, papers, and
ideas, measured almost entirely through publications and their citations.
Fortunato et al.~\citep{fortunato2018sciofsci} note explicitly that ``most SciSci
research focuses on publications as primary data sources'' and treat science as an
economic system with a ``one-dimensional currency of citation counts''; even landmark
results in this tradition (e.g.\ that large teams develop while small teams disrupt science
and technology~\citep{wu2019disruption}) are derived entirely from the paper-and-patent
citation graph, with software nowhere in the measurement;
software appears only once, in a passing list of ``scholarly artifacts,'' and
research tooling is otherwise treated as a cost driver of big science rather than
as a measurable output. Traag~\citep{traag2022citationmodels} confirms that the
entire apparatus of citation distributions, ageing models, and evaluation
indicators operates on the paper-to-paper graph, with no software channel, while
cautioning (via Goodhart's and Campbell's laws) that any indicator reshapes the
behavior it measures. From the software side, Afiaz et
al.~\citep{eval2023biomedimpact} foreground software as the object of evaluation
but their study is single-tool, developer-centric, and small-scale (48 surveyed
developers, 44 manually inspected tools in one funding program), built on noisy,
identity-unresolved text-mention data, and explicitly unable to measure what
matters most: the downstream use of a tool's outputs, how a tool seeds further
tools, and the scholarly value of the work it enables. Science-of-science has the
scholarly-outcome linkage and global scale but omits software; software-impact
evaluation centers software but lacks cross-corpus scale, identity resolution, and
a principled link to scholarly outcomes.

\subsection{A new class of contributor: AI agents in open-source software}
\label{sec:rel-aicensus}
A new and rapidly growing class of open-source contributor has begun to enter the
software supply chain: autonomous AI coding agents. Large language models are already
reshaping scientific production itself~\citep{kusumegi2025scientific}, and their reach now
extends from the writing of papers to the writing of the software those papers depend on. Khosravani and
Mockus~\citep{aicensus2026} census their prevalence across more than 180 million
Git repositories in World of Code over three snapshots spanning December 2024 to
April 2026, using a four-method detection framework (configuration-file scanning,
commit-message signatures, distributed author-identity matching, and centralized
bot-account lookup). Their central result is that no single signal is
representative: multi-method detection finds 850{,}157 Claude Code commits across
17{,}295 projects in the V2510 snapshot, of which the bot-account lookup most adoption
studies rely on recovers
only 28{,}154 (3.3\%), a 30$\times$ relative-recall gap, and a comparison against a
pull-request-based census shows the two channels capture nearly disjoint agent
populations (a PR census misses 79\% of commit-detected Claude Code adopters). By
the most recent (V2604) snapshot, commit-attributed agents generate over 320{,}000 commits
per month (Claude Code alone contributing 886{,}122 commits,
half of all AI-attributed commits), and agent activity rises from 1.6\%
to 6.7\% of non-bot commits within adopting projects in a single quarter. This
matters to our study in two ways. First, it stresses the same identity substrate
our measurement relies on: the census itself depends on World-of-Code author
aliasing~\citep{AZBZM19} and deforking to avoid inflating contributor and project
counts, and must deliberately keep agent identities \emph{separate} from human
alias clusters (our pipeline likewise separates \texttt{bot}/\texttt{bad}/
\texttt{local} from \texttt{developer} identities), so ``who builds scientific
software'' is becoming a moving target that only robust identity resolution can
track. Second, AI agents are themselves a science$\rightarrow$software channel
(research models deployed as tools that now \emph{produce} software at scale),
making the reciprocal science$\leftrightarrow$software loop we measure both larger
and harder to attribute, and raising provenance and licensing questions for the code such
agents emit and ingest~\citep{llmCuration}. We position our cross-corpus graph as the substrate
on which such agent contributions can be located, attributed, and folded into impact
accounting.

\subsection{Gap and contributions}
\label{sec:rel-gap}
Across these threads the same gap recurs: no prior work links scientific papers to
software artifacts at \emph{global, cross-ecosystem scale} with
\emph{resolved developer identity}, traversable \emph{reciprocally}. Mention
mining reaches the artifact only sparsely and unreliably, by bare name matching;
dependency analysis reaches the artifact
but only within a few managed ecosystems and only \emph{into} software;
science-of-science reaches scholarly outcomes but omits software entirely. We
close this gap with a cross-corpus graph over OpenAlex/Semantic Scholar and World
of Code that (1) links papers to repositories through both mention and
dependency layers, (2) resolves developers to canonical identities so authorship
and reciprocal credit are observable, and (3) measures impact in both
directions, how literature shapes scientific software (RQ1) and how software
enables research (RQ2), using SciCat's \nscicat{} science repositories as a
worked example. Throughout, we respect the metric-distortion and causal cautions
that Traag~\citep{traag2022citationmodels} and Afiaz et
al.~\citep{eval2023biomedimpact} raise: dependency in-degree and reciprocal links
are reported as \emph{complementary} evidence of importance, not as a single
gameable score.

\section{Data and the Cross-Corpus Graph}
\label{sec:data}

\paragraph{Corpora.}
\emph{World of Code} (WoC, version \texttt{V2604})~\citep{woc19} provides the
near-complete graph of public version-control history: commits, authors, projects,
blobs, and project-to-project dependency edges. \emph{Semantic Scholar}~\citep{kinney2023semanticscholar}
and \emph{OpenAlex}~\citep{openalex} provide the bibliographic side: papers, authors,
citations, affiliations, and software/repo mentions.

\paragraph{Seed: science repositories.}
We anchor on \textbf{18{,}247 science repositories} from the SciCat
dataset~\citep{malviyathakur2023scicatcurateddatasetscientific,scisust}, each carrying a
WoC \texttt{ProjectID}, a scientific Field, a 0--4 reuse-layer taxonomy, and popularity
metrics (stars/forks). SciCat is assembled by LLM-classifying World-of-Code projects for
scientific intent and curating the survivors; the same seed underpins a study finding
scientific OSS to be more sustainable than commonly assumed~\citep{scisust}.

\paragraph{Typed graph.}
The cross-corpus graph (\texttt{cite-study/graph/edges.typed.gz}) has 69{,}775{,}250
edges over eight layers (Table~\ref{tab:layers}). Two layers carry the reciprocal signal:
\textsc{mentions\_doi} (software cites a paper; channel A) and \textsc{mentions\_repo}
(a paper names software; channel B). \textsc{depends\_on} (49.2M project$\to$project edges)
carries software-on-software reuse. Identity is reconciled by \textsc{same\_as} bridges
(DOI co-occurrence, OpenAlex ORCID/affiliation, and repo-URL anchoring) and by the WoC
author-aliasing map~\citep{AZBZM19}.

\begin{table}[t]\centering\small
\caption{Layers of the cross-corpus graph (V2604), counts from
\texttt{edges.typed.gz}.}
\label{tab:layers}
\begin{tabular}{llr}
\toprule
Layer & Endpoints & Edges \\
\midrule
\textsc{depends\_on}      & project $\to$ project & 49{,}207{,}245 \\
\textsc{has\_author}      & paper $\to$ S2-author & 8{,}594{,}898 \\
\textsc{publishes}        & project $\to$ package & 7{,}096{,}118 \\
\textsc{maintained\_by}   & package $\to$ author  & 2{,}451{,}395 \\
\textsc{mentions\_doi}    & project $\to$ paper (chan A) & 2{,}401{,}620 \\
\textsc{mentions\_repo}   & paper $\to$ project (chan B) & 12{,}310 \\
\textsc{same\_as}         & cross-corpus identity & 10{,}143 \\
\textsc{affiliated\_with} & author $\to$ institution & 1{,}521 \\
\midrule
Total & & 69{,}775{,}250 \\
\bottomrule
\end{tabular}
\end{table}

\paragraph{Coverage of the seed.}
Of 18{,}247 science repositories: \textbf{6{,}852 (37.6\%)} cite or are named by $\ge$1
paper, \textbf{4{,}652 (25.5\%)} publish a package, and \textbf{2{,}025 (11.1\%)} are
depended upon by $\ge$1 other project. The join yields 63{,}240 (repo, DOI) pairs spanning
58{,}476 distinct papers and 45{,}764 (repo, scholarly-author) pairs over 39{,}278 authors.
Lowercasing the seed and graph join keys to the WoC canonical (\texttt{url2woc}) space
reconciles repository-name casing and recovers 8{,}115 uppercase science repositories that a
case-preserving join had silently dropped, a $+74\%$ gain on the dependency lens.

\section{Methods: Two Reciprocal Impact Measures}
\label{sec:methods}

We operationalize each RQ as an aggregation over directed graph paths. The two directions are
\emph{not} expressed in a common currency: \rqA{} reach is a count of distinct repositories,
\rqB{} ecosystem impact a count of distinct dependent projects, and the two run over different
edge types. We therefore treat them as complementary lenses on the same graph, comparing
\emph{which strata each surfaces} (\S\ref{sec:synthesis}), and do not net them into a single
balance or claim one magnitude exceeds the other.

\paragraph{\rqA{}: science $\to$ software.} For a paper $d$, its
\emph{software reach} $R_{\text{sw}}(d)$ is the number of distinct science repositories
that cite $d$ (reverse \textsc{mentions\_doi}). For a repository $p$, its
\emph{literature grounding} $G(p)$ is the number of distinct papers it cites. For a
scholarly author $s$, \emph{toolmaker reach} is the number of distinct science repositories
reachable via $s$'s papers (reverse \textsc{mentions\_doi} $\circ$ \textsc{has\_author}).

\paragraph{\rqB{}: software $\to$ science.} For a repository $p$, its
\emph{scientific uptake} is the number of papers naming it (\textsc{mentions\_repo}), and
its \emph{ecosystem impact} $I(p)$ is its reverse-\textsc{depends\_on} in-degree: the
number of distinct projects that depend on it. The unit of ``science'' for \rqB{} is the
scholarly paper (a DOI-identified work): \emph{scientific uptake} counts the distinct papers
that directly name a repository (a one-hop reverse \textsc{mentions\_repo} grounding), and
\emph{ecosystem impact} counts the distinct projects that directly depend on it (a one-hop
reverse \textsc{depends\_on}). We deliberately do \emph{not} follow transitive
software$\to$software$\to$paper or paper$\to$paper chains: each measure is the direct
grounding/dependency of a single repository, not a multi-hop science$\to$science closure, so
no transitive citation or dependency cascade is collapsed into the count.

\emph{Scope of the two \rqB{} measures, stated plainly.} \emph{Scientific uptake}
(\textsc{mentions\_repo}) is the only measure here that is software$\to$\emph{science} in the
strict sense: its target is a paper. Its denominator, however, turns out to be tiny and its
recall near zero against gold (\S\ref{sec:validation}), so it cannot by itself rank software by
scientific enabling. \emph{Ecosystem impact} (reverse \textsc{depends\_on}) is the measure that
actually populates Table~\ref{tab:sw-impact}, and its dependents are \emph{arbitrary
World-of-Code projects}: the entire software ecosystem, not the science seed. It is therefore a
software$\to$\emph{software} reuse measure that we use as a \emph{proxy} for a tool's scientific
enabling, on the premise that software depended upon at ecosystem scale is also the substrate of
much downstream science. We keep the two distinct throughout and do not claim the dependents are
themselves scientific; treating dependency in-degree as the software$\to$science answer would
overstate what \textsc{depends\_on} can see. Both measures are deliberately \emph{first-order}
(one-hop in-degree), not centrality-weighted: a Katz- or PageRank-style measure that weights a
dependent by its own importance is a natural refinement we leave to future work, noting that on
heavy-tailed dependency graphs raw and centrality-weighted in-degree are typically strongly
rank-correlated.

\paragraph{Hygiene.} DOIs are lowercased on both sides; placeholder/template DOIs
(\texttt{zenodo.xxxxxx}, \texttt{1234}, \texttt{doi}, \dots) are filtered; join keys are
mapped to the WoC canonical (\texttt{url2woc}) space. We assess robustness to the principal
linkage errors downstream: sensitivity to mention-extraction recall is reported in
\S\ref{sec:rqB} (thinning both counts to simulated recall $0.7$, $0.5$, and $0.3$) and bounded
against the Softcite gold corpus in \S\ref{sec:validation}, and \textsc{same\_as} precision is
treated as a threat in \S\ref{sec:discussion} (multi-evidence bridges, toolmaker-recovery as a
positive control).

\section{Results}
\label{sec:results}

We present the following as a set of basic feasibility investigations: each asks whether a
particular science-supply-chain quantity can be \emph{computed at all} from the cross-corpus
graph and whether the resulting signal is coherent, rather than offering a final or causal
accounting. Read together, they probe whether the conceptualization holds up enough to be worth
building out, and where the linkage must improve before stronger claims are warranted. We flag
the principal caveats inline and consolidate them in \S\ref{sec:validation} and
\S\ref{sec:discussion}.

\subsection{\rqA{}: The Impact of Science on Software}
\label{sec:rqA}

\paragraph{Which papers reach furthest into software.}
Table~\ref{tab:paper-reach} ranks papers by the number of distinct science repositories
that cite them. The top of the list is \emph{not} the most-cited science overall but the
\emph{reproducibility/packaging} layer (nf-core, MultiQC, Nextflow, Singularity,
Bioconda), infrastructure that pipelines adopt wholesale, followed by heavily used
sequence-analysis tools (SAMtools, STAR, Bowtie\,2, BWA). Broadest repository adoption
(nf-core, 55 repos) is distinct from highest scholarly citation count (SAMtools, 53{,}220
S2 cites).

This inversion is the science$\to$software analogue of the weak reuse--citation coupling we quantify in
\S\ref{sec:rqB}: what propagates most widely into research software is not the most
\emph{cited} science but the most \emph{operationally reusable} layer. A workflow framework or
container recipe is adopted wholesale by many pipelines (each a fresh repository) whereas a
heavily cited primary method (SAMtools, Bowtie\,2) is invoked through that shared packaging
layer and so accrues citations without itself multiplying across repositories. Repository reach
and citation count therefore measure different things: reach rewards being a substrate that
others build \emph{on}, citation rewards being a result others build \emph{from}. Reading
software impact off citation counts alone would systematically demote the packaging/workflow
tier that in fact carries the literature's broadest contact with software.

\paragraph{These rankings are a structurally-biased slice, not a complete reach measure.}
The \rqA{} rankings here are conditional on the mention surface our graph can observe: paper
abstracts, landing-page locations, and reference lists. Our held-out check against the Softcite
gold corpus (\S\ref{sec:validation}) shows recall is near zero on \emph{methods-section}
mentions, the surface where most working software is actually named. A tool invoked only in
methods prose (never in an abstract, a structured reference, or a registered landing page) is
therefore largely invisible to these counts. The ordering in Table~\ref{tab:paper-reach} and the
grounding ranking in Table~\ref{tab:lit-grounding} should accordingly be read as a thin,
structurally-biased cross-section of true software reach (one that over-represents tools with a
formal citable announcement and a packaging footprint and under-represents tools mentioned only
in passing in methods text) rather than as a complete or unbiased reach census.

\begin{table}[t]\centering\small
\caption{\rqA{}. Papers with the broadest reach into science software
(\# distinct science repos citing), with year and Semantic Scholar citation count.}
\label{tab:paper-reach}
\begin{tabular}{lrrr}
\toprule
Paper (software) & \#repos & year & S2 cites \\
\midrule
nf-core framework for community bioinformatics pipelines & 55 & 2020 & 2{,}297 \\
MultiQC: summarize results for multiple tools           & 55 & 2016 & 6{,}619 \\
Nextflow: reproducible computational workflows          & 50 & 2017 & 2{,}829 \\
Singularity: scientific containers                      & 50 & 2017 & 1{,}954 \\
Bioconda: sustainable software distribution             & 47 & 2018 & 1{,}322 \\
BioContainers                                           & 43 & 2017 & 694 \\
SAMtools                                                & 29 & 2009 & 53{,}220 \\
STAR: ultrafast RNA-seq aligner                         & 14 & 2013 & 11{,}804 \\
Bowtie\,2                                               & 12 & 2012 & 46{,}860 \\
Matplotlib                                              & 10 & 2007 & 21{,}776 \\
\bottomrule
\end{tabular}
\end{table}

\paragraph{Which repositories are most grounded in the literature.}
A naive ranking by distinct papers cited is dominated by \emph{bibliography/data
catalogues}: repositories whose content \emph{is} an enumeration of study DOIs (a curated
RNA-seq study list, a forest-carbon database) rather than software grounded in the methods it
implements. We separate these with a reproducible rule: a repository is a catalogue if it
cites at least 300 distinct DOIs, at least 95\% of which no other science repository cites
(a bibliography unique to it), and its DOI count is at least twice its commit count, i.e.\
references dwarf development activity (where commit metadata is missing, the singleton-DOI
test alone applies). This flags only \textbf{5 repositories}
(geo-rnaseq 9{,}395; preta 4{,}952; \texttt{igbb\_past} 2{,}335; \texttt{forc-db} 1{,}443;
\texttt{ithim-r} 495) and is deliberately conservative: meta-analysis and simulation tools
that legitimately reference many studies (metafor, NEURON, HyPhy) are retained because their
development activity is commensurate with their bibliography. Table~\ref{tab:lit-grounding}
ranks the surviving \emph{software}; genuine reusable tools cluster at the hundreds level.

\begin{table}[t]\centering\small
\caption{\rqA{}. Software most grounded in the literature (\# distinct papers cited), after
removing 5 bibliography/data catalogues (see text).}
\label{tab:lit-grounding}
\begin{tabular}{lrll}
\toprule
Repository & \#papers & Field & layer \\
\midrule
xrobin\_proc (pROC) & 1{,}739 & Statistics & Domain-specific \\
wviechtb\_metafor & 1{,}205 & Statistics & Domain-specific \\
rcastelo\_gsva (GSVA) & 892 & Biology & Domain-specific \\
neuronsimulator\_nrn (NEURON) & 765 & Neuroscience & Domain-specific \\
veg\_hyphy (HyPhy) & 721 & Biology & Domain-specific \\
rglab\_mast (MAST) & 667 & Biology & Domain-specific \\
geodynamics\_specfem3d\_globe & 599 & Earth Science & Domain-specific \\
grimbough\_biomart (biomaRt) & 445 & Biology & Domain-specific \\
\bottomrule
\end{tabular}
\end{table}

\paragraph{Which scholars underpin research software.}
Ranking scholarly authors by the number of science repositories reachable through their
papers surfaces \emph{toolmakers}, not random homonyms: Paolo Di Tommaso (Nextflow, reach
63), G.~M.~Kurtzer (Singularity, 51), S.~Salzberg (Bowtie/Salmon, 23), Heng Li
(SAMtools/BWA/minimap2, 12). These four are illustrative of the head of the ranking rather than
a systematic audit; that the highest-reach scholarly authors are nonetheless precisely the
canonical developers of the tools their papers announce is a positive control on the
DOI-anchored author linking: a chain that mis-resolved DOIs, mis-merged author identities, or
mis-attributed repositories would surface homonyms and unrelated names rather than the actual
toolmakers, so this clean recovery is direct evidence that the \textsc{has\_author},
\textsc{mentions\_doi}, and \textsc{same\_as} bridges compose coherently. The underlying
developer-identity resolution is the production WoC aliasing map~\citep{AZBZM19}, whose
construction and validation at global scale is detailed in a companion
paper~\citep{aliasingcompanion}.

\subsection{\rqB{}: The Impact of Software on Science}
\label{sec:rqB}

\paragraph{The hidden-infrastructure tier.}
Ranking science repositories by reverse-\textsc{depends\_on} in-degree
(Table~\ref{tab:sw-impact}) surfaces a \emph{machine-learning / data-science infrastructure}
tier: PyTorch (12{,}864 dependents), seaborn, NLTK, einops, TF.js, Hydra, and CatBoost. This
tier is largely \emph{disjoint} from the bioinformatics tools that dominate \rqA{}. These are the
substrates on which downstream science is built, yet they are rarely cited as such. As stated in
\S\ref{sec:methods}, the dependents counted here are general World-of-Code projects, not science
repositories: this ranking is a software$\to$software reuse proxy for scientific enabling, used
because the strict software$\to$science channel (\textsc{mentions\_repo}) is too sparse to rank
(\S\ref{sec:validation}). The 12{,}864 dependents of PyTorch are the ecosystem at large; the
claim is that a tool the ecosystem leans on this heavily is also a load-bearing substrate for the
science built atop that ecosystem, not that all 12{,}864 are themselves scientific.

\begin{table}[t]\centering\small
\caption{\rqB{}. Science software by ecosystem impact (reverse \textsc{depends\_on}
in-degree). Dependents and stars are only weakly coupled.}
\label{tab:sw-impact}
\begin{tabular}{lrrr}
\toprule
Repository & \#dependents & Field & stars \\
\midrule
pytorch\_pytorch     & 12{,}864 & CS & 56{,}634 \\
cloudflare\_circl    & 9{,}459  & CS & 150 \\
mwaskom\_seaborn     & 6{,}563  & Data Sci & 7{,}546 \\
nltk\_nltk           & 2{,}006  & CS & 17{,}033 \\
arogozhnikov\_einops & 1{,}567  & CS & 2{,}381 \\
imageio\_imageio     & 1{,}207  & CS & 784 \\
pytorch\_audio       & 1{,}144  & CS & 1{,}033 \\
tensorflow\_tfjs     & 790      & CS & 25{,}422 \\
napari\_napari       & 573      & Biology & 514 \\
\bottomrule
\end{tabular}
\end{table}

\paragraph{Reuse is not popularity.}
Dependency in-degree and GitHub stars are only loosely coupled: \texttt{cloudflare\_circl}
has 150 stars but 9{,}459 dependents; \texttt{einops} has 2{,}381 stars and 1{,}567
dependents. Across all 2{,}025 depended-upon science repositories the rank correlation
between dependency in-degree and stars is only moderate (Spearman
$\rho=0.36$, bootstrap 95\% CI $[0.32,0.40]$; Pearson on $\log$-counts $r=0.38$), and weaker
still against forks ($\rho=0.33$, bootstrap 95\% CI $[0.29,0.37]$); the coupling is
near-identical in natural-science and
computer-science repositories ($\rho=0.31$, $[0.25,0.37]$ vs.\ $0.34$, $[0.28,0.40]$; CIs
overlap), so it is not a between-field artifact.
Stars therefore explain little of the variance in actual reuse.
The reuse--citation relationship is weaker than reuse--stars, but its strength and even its
sign depend on how a repository is paired with a citation count, and we report that sensitivity
rather than a single estimate (Table~\ref{tab:reusecite}). Through the \emph{naming} channel,
for the 137 depended-upon repos that at least one paper names in our graph, dependency
in-degree is statistically \emph{uncorrelated} with the citation count of that paper (Spearman
$\rho=0.05$, $p=0.55$, bootstrap CI $[-0.11,0.20]$): a null we cannot reject. But the naming
channel is the sparsest and most selection-biased link in the graph, over-representing
methods/tool papers that name their software in prose. Pairing each repository instead with the
DOIs it \emph{declares} for itself (\textsc{citation.cff}, Zenodo badges) widens the support
almost eightfold, to $n=1{,}067$, and the correlation becomes a weak but resolvable positive:
$\rho=0.13$, bootstrap CI $[0.07,0.19]$, now excluding zero. The shift survives a field split:
in the larger panel both strata couple positively (natural science $\rho=0.11$, CI
$[0.03,0.19]$, $n=645$; computer science $\rho=0.21$, CI $[0.12,0.31]$, $n=422$), whereas in
the naming panel both straddled zero. We read the gap between the two panels, not either number
alone, as the result: at the linkage density science--software graphs currently support, the
sign of the headline reuse--citation correlation is an artifact of which coverage gap one
tolerates, so any strong decoupling claim is premature. The larger panel is at least not a
random-missingness artifact: thinning both counts to recall $0.7$, $0.5$, and $0.3$ leaves
$\rho$ near $0.12$ rather than collapsing toward zero. That thinning is \emph{random}
(missing-completely-at-random); it bounds uniform under-counting but not the \emph{structured}
surface bias the gold benchmark reveals, where methods-section mentions are dropped
systematically (\S\ref{sec:validation}), so we cannot rule out that a recall mechanism
correlated with citation count would move either estimate further.

\begin{table}[t]\centering\small
\caption{Reuse versus citation rank correlation under two pairings of a repository with a
citation count: the naming panel (papers that name a repository) and the declared-DOI panel
(DOIs a repository mints for itself).}
\label{tab:reusecite}
\begin{tabular}{lrll}
\toprule
Pairing channel & $n$ & $\rho$ & 95\% CI \\
\midrule
Paper names repository      & 137   & $0.05$ & $[-0.11,\ 0.20]$ \\
Repository declares DOI     & 1{,}067 & $0.13$ & $[\phantom{-}0.07,\ 0.19]$ \\
\quad natural science       & 645   & $0.11$ & $[\phantom{-}0.03,\ 0.19]$ \\
\quad computer science      & 422   & $0.21$ & $[\phantom{-}0.12,\ 0.31]$ \\
\bottomrule
\end{tabular}
\end{table}

\paragraph{Statistical reporting.} All coupling claims above are stated as rank (Spearman) or
log-scale correlations with bootstrap confidence intervals; because the underlying dependency,
star, and citation counts are heavy-tailed, a raw-scale (linear Pearson) correlation is
dominated by a few extreme repositories and understates the association; we therefore rely on
the rank/log-log estimates and do not report or interpret the skew-distorted linear value. We
apply no multiple-comparison correction across the several decoupling and overlap tests; the
conclusions we draw are based on the direction and confidence intervals of effects (which
straddle or exclude zero) rather than on any single thresholded $p$-value, and should be read
accordingly.

\subsection{Synthesis: The Two Directions Surface Different Strata}
\label{sec:synthesis}

\begin{table}[t]\centering\small
\caption{Three complementary impact lenses over one graph.}
\label{tab:lenses}
\begin{tabular}{lll}
\toprule
Lens & Edge & Stratum surfaced \\
\midrule
\rqA{} paper$\to$software & reverse \textsc{mentions\_doi} & bioinformatics reproducibility/packaging \\
\rqA{} repo$\to$literature & \textsc{mentions\_doi} count & data-aggregation + stats/bio tools \\
\rqB{} software impact & reverse \textsc{depends\_on} & ML/DS infrastructure \\
\bottomrule
\end{tabular}
\end{table}

\begin{figure}[t]
\centering
\includegraphics[width=0.92\linewidth]{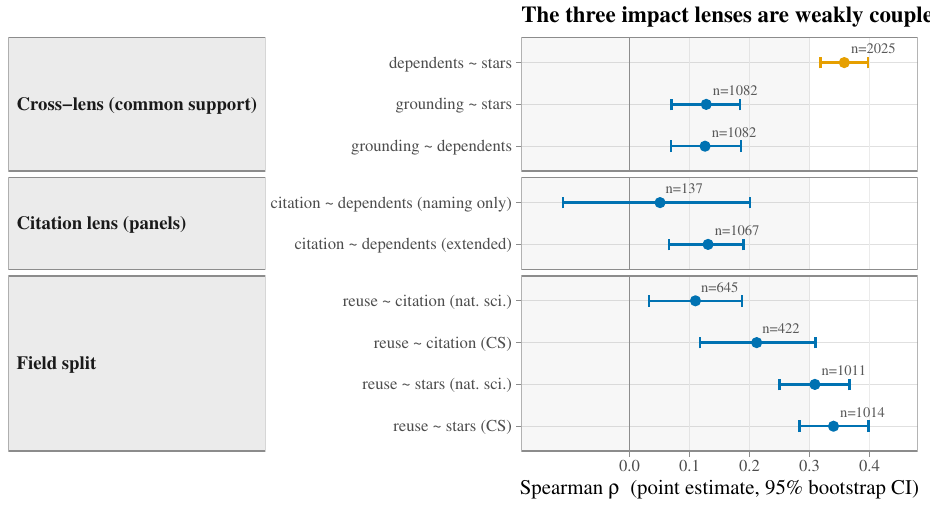}
\caption{Rank associations across the three impact lenses, each a Spearman
$\rho$ with a 95\% bootstrap CI over the lens's common support. Point estimates
cluster near zero and CIs mostly fall short of moderate correlation
($\rho\!\ge\!0.3$, shaded), the evidence that the lenses are weakly coupled.}
\label{fig:decoupling}
\end{figure}

Across three lenses, broad ecosystem reuse, scholarly citation, and popularity surface largely
distinct strata (Figure~\ref{fig:decoupling}); stars do not substitute for dependency reuse.
The same graph, read in three directions, yields three \emph{weakly coupled} pictures of
impact (Table~\ref{tab:lenses}) that are far from interchangeable; none is recoverable from
scholarly citation count alone. Over each lens's common support the rank correlations are
modest: grounding vs.\ dependency $\rho=0.13$ (95\% CI $[0.07,0.19]$), grounding vs.\ stars
$\rho=0.13$ $[0.07,0.18]$, dependency vs.\ stars $\rho=0.36$ $[0.32,0.40]$. A repository's
position on one axis predicts little of its position on another. The top-ranked sets reflect
the same story but must be read against the right null: with thousands of candidate
repositories, two random top-50 lists would overlap by chance at Jaccard $\approx0.003$, so a
\emph{near-zero} Jaccard is not by itself informative. We therefore compare the observed
overlap to a 2{,}000-draw random-subset null. At $k=50$, grounding vs.\ dependency overlaps at
$J=0.042$ (4 of 50 shared; $14\times$ the chance level, permutation $p=0.001$) and dependency
vs.\ stars at $J=0.111$ ($9\times$ chance, $p<0.001$), both well \emph{above}
independence in magnitude (permutation $p\le0.001$), evidence of a shared-but-weak structure, whereas grounding vs.\ stars
($J=0.010$) is indistinguishable from chance ($p=0.27$). The three-way intersection of the
top-50 sets contains a single repository (\texttt{mwaskom\_seaborn}). Each lens thus surfaces a
largely distinct stratum of the science--software ecosystem: the lenses are correlated enough
to share a faint common signal, but nowhere near enough for any one, least of all popularity
(stars), to substitute for another.

\subsection{What Is Associated with AI Coding-Agent Adoption}
\label{sec:aiagents}

AI coding agents are a 2023+ phenomenon, whereas the reuse and citation stocks studied above
accumulated over the preceding decade. Regressing those stocks on adoption would invert the
arrow of time. Any long-run effects of delegating scientific software work to AI agents will
take years to surface in dependency and citation counts; in the interim we ask
the identifiable mirror-image question: \emph{what kind of science repository adopts AI agents
first?} We detect agent activity at the commit level over the World of Code object store
(configuration-file, commit-message, and bot-account signatures), yielding 332{,}693 distinct
AI-agent commits universe-wide. This 332{,}693-commit set is the V2604-intersected, all-agent
detection used here; it is not directly comparable to the companion AI census's $850{,}157$
Claude-Code commits~\citep{aicensus2026}, which counts a single agent in the earlier \texttt{V2510}
snapshot of the full World of Code object store rather than the all-agent, V2604-restricted universe we join
against the SciCat seed. Mapping our set to projects (\textsc{c2p}) and intersecting with
the seed identifies \textbf{58 adopting science repositories} (0.32\%), including mainstream
computational-science tools (\texttt{snakemake}, \texttt{nextflow}, \texttt{deepmd-kit},
\texttt{wfmash}, \texttt{matgl}).

\paragraph{Adoption is only weakly forecastable from antecedents.} We frame adoption as a
held-out predictive task: logistic regression with stratified 5-fold cross-validation
(averaged over five seeds), scored by ROC-AUC and rare-positive PR-AUC. Features fixed
\emph{before} the AI era (project age, scientific field, and software-reuse layer) yield only
AUC $=0.60\pm0.01$ (chance $=0.50$; Table~\ref{tab:aiadopt}): which projects adopt is largely unpredictable
from their pre-existing structural profile. Adding contemporaneous development intensity
($\log$ commits, authors, active months, stars, forks, files) raises this to AUC
$=0.74\pm0.01$ (PR-AUC $0.038$, a $12\times$ lift over the $0.32\%$ base rate), but those
covariates are measured at the watermark and so \emph{describe} the adopter rather than predict
it from antecedents.

\begin{figure}[t]
\centering
\includegraphics[width=0.88\linewidth]{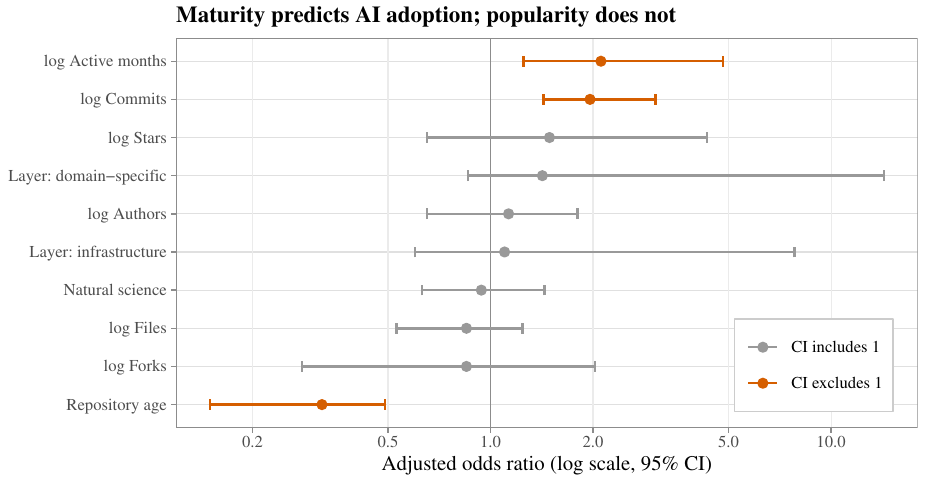}
\caption{Adjusted odds ratios (95\% CI, log scale) for AI-toolchain adoption.
Maturity and sustained activity move the odds; popularity (stars, forks) and
field do not, with CIs straddling the no-effect line at $1$.}
\label{fig:facilitator-or}
\end{figure}

\paragraph{Adoption is associated with development intensity and relative youth.} In a penalized
(L2) logistic model with bootstrap CIs (Table~\ref{tab:aiadopt}, Figure~\ref{fig:facilitator-or}), the factors independently
associated with higher adoption odds are sustained activity (active months, standardized OR $2.11$, 95\% CI
$[1.25,4.82]$; commit volume, OR $1.96$, $[1.43,3.05]$) and, conditional on that activity,
relative \emph{youth}: among equally active projects the more recently founded adopt sooner
(age OR $0.32$, $[0.15,0.49]$). Team size, popularity, file count, and the natural-science-vs-CS
contrast add nothing once intensity is controlled (all CIs span $1$). Raw rates echo this:
adoption peaks in fast-moving computational fields (Quantum Computing 1.7\%, 3 of 172;
Chemistry 0.9\%, 5 of 561; Data Science 0.6\%, 5 of 874) and is zero in Mathematics (0 of 437),
Statistics (0 of 347), and Neuroscience (0 of 324), but field washes out under the intensity
control. These per-field rates rest on single-digit adopter counts and are reported only as
descriptive context, not as field-level effects. AI adoption is, so far, a behavior of active, fast-moving
software projects rather than of any particular science.

\begin{table}[t]\centering\small
\caption{Factors associated with AI-agent adoption among the \nscicat{} seed repositories ($58$
adopters, $0.32\%$). \emph{Top:} held-out predictive performance; \emph{bottom:} standardized
odds ratios with bootstrap 95\% CIs ($\ast$ marks a CI excluding $1$).}
\label{tab:aiadopt}
\begin{tabular}{lrr}
\toprule
\emph{Predictive task} & ROC-AUC & PR-AUC \\
\midrule
Antecedent only (age, field, layer)      & $0.60$ & $0.005$ \\
\quad $+$ contemporaneous size            & $0.74$ & $0.038$ \\
\midrule
\emph{Associated factor} & \multicolumn{2}{c}{std.\ OR\ [95\% CI]} \\
\midrule
Active months              & \multicolumn{2}{c}{$2.11\ [1.25,4.82]\,\ast$} \\
Commit volume              & \multicolumn{2}{c}{$1.96\ [1.43,3.05]\,\ast$} \\
Project age (older)        & \multicolumn{2}{c}{$0.32\ [0.15,0.49]\,\ast$} \\
Stars                      & \multicolumn{2}{c}{$1.49\ [0.65,4.32]$} \\
Team size / files / nat.\ sci. & \multicolumn{2}{c}{n.s.\ (CI spans $1$)} \\
\bottomrule
\end{tabular}
\end{table}

\paragraph{Why not (yet) a post-adoption outcome study.} We also attempted the post-adoption
design directly. Of the 58 adopters, 15 have AI commits present in the V2604 object store with
recoverable timestamps; their first AI commit falls between 2024-02 and 2025-04 (median
2024-10), leaving a median follow-up of just \textbf{1.4 years} (IQR $1.2$--$1.8$, max $2.1$)
to the watermark. Because our outcomes are cumulative \emph{stocks} read at a single
snapshot (not a pre/post panel) a within-repository difference-in-differences is not
constructible, and a one-to-two-year window is far too short for new dependents or citing
papers to accrue. We therefore report adoption antecedents now and flag the outcome study as a
deliberate follow-up, identifiable once a second watermark snapshot and a longer exposure
window exist. We defer the pre/post outcome panel to a future World of Code snapshot, which will
supply the second watermark and the additional exposure time the difference-in-differences
design requires.

What the cross-corpus graph contributes is the ability to pose either question at all: to
follow a new mode of software production (AI-assisted commits) into the S3C
rather than stopping at the repository boundary. Commit-level detection has finite recall and
the treated group is modest, so we make no causal claim in either direction; the antecedent
model is descriptive, and the outcome model is deferred by construction.

\subsection{How Complete Is the Linkage? Internal and External Validation}
\label{sec:validation}

Every result above rests on \textsc{mentions\_doi}/\textsc{mentions\_repo} edges that are
\emph{manifest}: a paper--software link is recorded only when one corpus's text names the
other. Prior software-mention work shows recall well below
one~\citep{istrate2022czidataset,du2021softcite}, so the graph is a lower bound and our
overlap and correlation estimates are conservative. Two checks bound the gap.

\paragraph{Internal: independent extractors are complementary, not redundant.} Our
paper$\to$repo edges come from two methodologically independent extractors: Semantic Scholar
full-text citation contexts (\textsc{s2orc}: 822 DOIs, 659 repos) and OpenAlex
abstracts/locations (10{,}607 DOIs, 9{,}162 repos). The two extractors share \emph{zero} DOIs and
only $28$ repositories ($4.2\%$ of the \textsc{s2orc} set): full-text and abstract mentions
surface almost entirely disjoint paper--software pairs (per-extractor source tags and the
overlap are recoverable directly from the released \texttt{graph/edges.paper\_repo.gz}, built by
\texttt{bin/compose\_paper\_repo\_layer.sh}). This directly confirms the
single-extractor recall problem (each extractor alone misses most links) and means our measured
grounding counts are conservative lower bounds. It also explains why the reuse--citation
correlation is so sensitive to which pairing channel we use (\S\ref{sec:rqB}): when each channel
recovers only a thin, non-overlapping slice of the true links, the channel one happens to pick
can set the sign of the headline correlation.

\paragraph{External: a held-out gold benchmark.} We validate against the Softcite v2 gold
corpus~\citep{du2021softcite}: 4{,}971 biomedical and economics articles with $5{,}093$
human-annotated software mentions ($1{,}831$ distinct tools), reconciled by curators and
entirely independent of our extractors. Three findings, each with an honest denominator,
emerge.
First, a \emph{hard scope ceiling}: $19.6\%$ of gold mentions name proprietary tools (SPSS,
SAS, Stata, MATLAB, Excel, GraphPad Prism) that have no public git repository by construction
and thus can never appear in WoC. Any git-corpus linkage (ours or anyone's) is bounded above
by the FOSS fraction, here $80.4\%$. Second, for that FOSS subset the \emph{targets do exist} in
WoC: $315$ gold mentions across $80$ distinct tool names (BLAST, MAFFT, ImageJ, Ensembl, \dots)
resolve by name to at least one WoC project, so a miss is a missing \emph{edge}, not a missing
node. Third, \emph{recall is near zero on full-text mentions}: of the $147$ gold
papers that also sit in our OpenAlex-scanned universe, $65$ are annotated as mentioning $\geq 1$
FOSS tool, yet our manifest graph links \emph{none} of them ($0/65$) to the corresponding repo;
the gold papers and our linked corpus share \emph{zero} DOIs outright. The reason is structural:
gold annotations come from methods-section prose, whereas our edges come from abstracts,
landing-page locations, and reference lists: almost disjoint surfaces. This is the strongest
possible form of the missingness premise: an independent, human-curated benchmark confirms that
manifest links recover only a thin, structurally-distinct slice of true paper$\to$software
relationships, so the true recall behind \S\ref{sec:rqB}'s mention counts is very small. We do
not attempt a disattenuation correction (the recall is too poorly characterized to invert);
instead, \S\ref{sec:rqB} reports the reuse--citation correlation across both pairing channels
and under \emph{simulated further thinning} of both counts, and the present benchmark, by showing
how thin and structurally biased each channel is, establishes why a single estimate would be
fragile and why we report the panel gap rather than one number. It also sharpens
the interpretation of every count in \S\ref{sec:rqA}--\S\ref{sec:rqB} as a conservative lower
bound.
A natural extension, left to future work, is to complement this human-curated biomedical gold
point with the larger machine-extracted CZI Software Mentions
corpus~\citep{istrate2022czidataset}, which would add a domain-matched (and CS-inclusive) recall
estimate at the cost of the gold corpus's curator-level precision.

\paragraph{Seed representativeness.} The seed is a curated subset, not a sample, and two checks
quantify the gap. First, against the Softcite-curated tool set: of the $1{,}586$ non-proprietary
tool names, $952$ have a matching-named project somewhere in WoC, but only $108$ sit in the
SciCat seed, so of the curator-annotated tools that exist in WoC the seed covers $11.3\%$
($9.4\%$ on the distinctive-name subset that strips generic namesakes). Weighting by citation
frequency, only $6$ of the $30$ most-mentioned FOSS tools present in WoC are in the seed; the
absentees are established, distinctively-named tools (Ensembl, MAFFT, HMMER, Clustal~W, Limma)
the classifier did not surface. Second, against a differently-constructed anchor, the $3{,}595$
repositories behind every accepted Journal of Open Source Software (JOSS) paper, each carrying a
DOI and a repository by construction: $90.7\%$ are present in WoC, yet only $11.0\%$ ($397$) are
in the SciCat seed, and none of the Softcite-missing established tools appear among them (JOSS
began in 2016, after those tools were published). The two operationalizations of ``scientific
software'' thus overlap by roughly one part in nine and neither recovers the other's omissions:
the construct is operationalization-dependent, so every count here is a lower bound and the field
mix reflects the seed rather than a census. The response is to triangulate across independent
anchors rather than privilege one seed; conditioning a seed on the existence of a paper link, by
contrast, would trade this lower-bound bias for survivorship bias.

\paragraph{Triangulating anchors.} We take a first step: union SciCat with the JOSS
repositories and a Softcite tool-name anchor (curator tool names basename-matched to WoC,
namesake-heavy keys dropped), and compute each impact channel per anchor and on the union of
$23{,}299$ repositories (Table~\ref{tab:triangulate}). Three things follow. First, the anchors
are complementary, not redundant: JOSS is the better-connected seed (about twice SciCat's
dependency participation, $23.4\%$ vs $11.1\%$, and $2.5\times$ its package-publishing rate),
answering ``would a better-connected anchor help'' in the affirmative on the ecosystem axis,
while grounding is comparable and scientific uptake stays thin ($\sim$$3.7\%$) for every anchor.
Second, the two curated anchors surface different impact strata yet agree where it matters: of
the union's fifty highest ecosystem-impact repositories, $36$ come from SciCat (the megatools:
PyTorch, seaborn, NLTK) and $15$ from JOSS (well-integrated research packages SciCat missed,
e.g.\ Hypothesis, Pooch), with seaborn independently surfaced by both, an agreement signal a
single seed cannot produce. Third, the basename-expanded Softcite anchor is near-inert
($0.9\%$ depended-upon), confirming that curator tool \emph{names} belong in the validation
layer, not the seed. A fourth anchor, the $29{,}051$ repositories publishing a CRAN or conda
package (\textsc{SciPkg}, a science-leaning registry), completes a gradient: it is the most
ecosystem-connected ($52\%$ depended-upon) but the least paper-grounded ($21\%$ cite a paper,
$1.2\%$ named by one), so a registry seed captures software \emph{infrastructure} more than
science-grounded software, whereas SciCat is paper-grounded but mid-connected.

Re-running the reuse--citation decoupling per anchor closes the loop. For each anchor we
correlate a repository's ecosystem impact (reverse-\textsc{depends\_on} in-degree) with the
citations of its announcing paper (its own paper for JOSS, its best-cited declared DOI
otherwise), over repositories with both quantities positive. The Spearman coefficient is
\emph{positive on every anchor} (SciCat $+0.13$, $n{=}981$; JOSS $+0.35$, $n{=}473$; SciPkg
$+0.14$, $n{=}3{,}467$; union $+0.17$, $n{=}4{,}205$): a weak-to-moderate coupling whose sign is
stable across independently-constructed seeds, strongest for JOSS because its paper--repository
link is direct. Where the single-channel analysis (\S\ref{sec:rqB}) found the coupling's sign
could flip with the pairing channel, triangulating over four anchors and pinning the announcing
paper yields a consistently positive, modest coupling; the non-JOSS magnitudes are lower bounds,
since they use a repo-declared-DOI proxy for the announcement.

\begin{table}[t]\centering\small
\caption{Per-anchor connectivity into the cross-corpus graph: share with $\ge1$ edge on each
channel, and reverse-\textsc{depends\_on} in-degree (ecosystem impact) among the depended-upon
(median\,/\,p90). Softcite is a basename-expanded tool-name set, shown as a negative control.}
\label{tab:triangulate}
\begin{tabular}{lrrrrrr}
\toprule
Anchor & $n$ & \%dep & \%grnd & \%uptk & \%pub & dep m/p90 \\
\midrule
SciCat   & 18{,}247 & 11.1 & 37.0 & 3.6 & 25.5 & 3/37 \\
JOSS     &  3{,}260 & 23.4 & 36.1 & 3.8 & 65.2 & 3/28 \\
Softcite &  2{,}195 &  0.9 &  1.6 & 0.2 &  3.2 & 2/32 \\
SciPkg   & 29{,}051 & \textbf{52.0} & 21.0 & 1.2 & 100.0 & 5/123 \\
Union    & 50{,}835 & 33.1 & 25.1 & 2.0 & 67.1 & 4/105 \\
\bottomrule
\end{tabular}
\end{table}

\section{Discussion}
\label{sec:discussion}

\paragraph{Implications.}
\emph{(i) For the science-of-science agenda.} Software is a measurable, dependency-grounded
third output alongside papers and patents. Unlike a citation, a dependency edge is a
\emph{revealed} act of reuse rather than a rhetorical one. Models of how ideas and credit flow
through science are incomplete while the software term is missing; the cross-corpus graph
supplies it at population scale, and our lens-disjointness result
(\S\ref{sec:synthesis}) warns that any single proxy substituted for it will sample only one
stratum. \emph{(ii) For research-software valuation and funding.} Citation- and star-based
metrics systematically miss the hidden-infrastructure tier (a cryptography library with 150
stars and 9{,}459 dependents); dependency in-degree is a sharper signal of which artifacts a
funder cannot afford to let rot, and a less easily gamed one: a star or a citation can be
solicited, but a \textsc{depends\_on} edge is a revealed act of reuse that another project's
build actually exercises (though it, too, can be inflated by trivial or vendored dependencies;
\S\ref{sec:discussion}, Table~\ref{tab:threats}). \emph{(iii) For reproducibility and supply-
chain risk.} The packaging/workflow layer (nf-core, Nextflow, Singularity, Bioconda) is the
literature's primary point of contact with software, which makes it both the highest-value
target for reproducibility investment and the largest blast radius for a single point of
failure. \emph{(iv) For the AI transition.} Because the graph spans contributors as well as
artifacts, it can track a new mode of production (AI coding agents) through to scientific
outcome (\S\ref{sec:aiagents}); whatever the causal story, the instrument to monitor that
transition as it scales now exists.

\paragraph{How far does feasibility actually reach?}
We frame these investigations as feasibility evidence, but the same evidence shows where
feasibility runs out, and the gaps are not at the margins. Of the $18{,}247$ seed repositories
only $2{,}025$ ($11\%$) carry any measurable dependency reuse, so the \rqB{} ranking speaks for
roughly one repository in nine. The literature-side pairing is thinner still: only $137$ of those
depended-upon repositories are named by a paper we can resolve, and the held-out gold benchmark
links \emph{none} of its $65$ in-scope cases (\S\ref{sec:validation}). The clearest symptom is
that the headline reuse--citation correlation changes sign and confidence when we swap one
sparse pairing channel for a wider one (Table~\ref{tab:reusecite}): a measurement whose
conclusion turns on which gap the analyst tolerates cannot yet support strong claims about how
software reuse relates to scholarly citation. We therefore present the construction as
attainable in principle, while reporting honestly that at current linkage density the
quantitative couplings it yields are fragile and the safer reading is directional, not
point-estimated.

\paragraph{Threats to validity.}
Table~\ref{tab:threats} catalogues the principal threats and the mitigations in place. The
recurring theme is that every cross-corpus edge is the product of an imperfect linkage, so our
claims are deliberately structural (orthogonality of lenses, weak reuse--citation coupling) and
ordinal (rankings, odds ratios) rather than point estimates of absolute magnitude.

\begin{table}[t]\centering\small
\caption{Threats to validity and mitigations.}
\label{tab:threats}
\begin{tabular}{p{0.20\linewidth}p{0.40\linewidth}p{0.32\linewidth}}
\toprule
Threat & Description & Mitigation / residual risk \\
\midrule
Construct: ``science software'' & The SciCat seed is one LLM-classified operationalization
from a sampled crawl; flagship repositories can be absent (e.g.\ the E3SM model; only an I/O
component is present). & Seed is curated and field-balanced; results are framed relative to the
seed, not as universe-wide census; a coverage check (\S\ref{sec:validation}) finds the seed
holds $11.3\%$ of curator-annotated tools that exist in WoC. Residual: coverage is a lower bound. \\
Internal: mention recall (channel B) & Paper$\to$repo mentions are extracted from text and miss
unlinked or paraphrased references. & Treated as a lower bound; the reciprocal channel A
(\textsc{mentions\_doi}) and \textsc{depends\_on} do not depend on it. \\
Internal: \textsc{same\_as} precision & DOI-, ORCID-, and URL-anchored identity bridges have
finite precision; a wrong bridge mislinks paper and code. & Bridges are multi-evidence;
toolmaker-recovery (\S\ref{sec:results}) is a positive control. Residual: tail false bridges. \\
Internal: author/identity aliasing & Cross-corpus author linking inherits WoC alias-map error
(over/under-merging). & Uses the production aliasing map~\citep{AZBZM19}; affects only the
toolmaker-reach measure (\S\ref{sec:rqA}), not the repo- or paper-level counts. \\
Grounding inflation & Bibliography/data catalogues enumerate DOIs as content. & Conservative
filter removes 5 unambiguous cases (\S\ref{sec:results}); borderline embedded-bibliography
repos remain and are flagged. \\
Dependency semantics \& validation asymmetry & \textsc{depends\_on} counts declared
dependencies, not runtime use or intensity; vendored/copied code is under-counted. Unlike the
mention channels, the dependency lens (which carries the entire \rqB{} ranking) has \emph{no}
external gold benchmark validating it. & In-degree is interpreted ordinally; copy-based reuse is a
known separate channel~\citep{vuln22}. The \rqB{} ranking is a software$\to$software reuse proxy
(\S\ref{sec:methods},~\S\ref{sec:rqB}), reported as such; an external dependency-edge gold set is
left to future work. \\
AI temporal ordering & Agent adoption (2023+) post-dates the decade-long reuse/citation
stocks, so an adoption$\to$output regression would invert time. & Reframed to adoption
\emph{antecedents} (held-out prediction, \S\ref{sec:aiagents}); the outcome study is deferred
to a second snapshot (median post-adoption window only $1.4$\,yr). \\
AI detection \& confounding & Commit-level agent detection has finite recall; adopters
($n=58$) are few. & Predictive/descriptive framing only, bootstrap CIs; no causal claim in
either direction. \\
External: time \& platform & A single WoC watermark (V2604) and the GitHub-dominated forge
mix. & Cross-forge coverage via WoC; temporal generalization untested. \\
\bottomrule
\end{tabular}
\end{table}

\section{Conclusion}
\label{sec:conclusion}
Treating scientific production as an S3C of co-equal papers and software, and attempting
to build the bill of materials that links them, begins to turn ``the impact of software on
science'' from a slogan into something measurable, and makes the gaps in
that measurement visible. Our investigations are a feasibility probe, not a final accounting: read
in three directions, one global cross-corpus graph appears to surface three weakly coupled strata
of impact (reproducibility infrastructure, hidden ML/data-science infrastructure, and the
toolmakers behind both), none of which is recoverable from scholarly citation counts or repository
popularity. In our data the coupling between dependency-based reuse and the citations of the paper
that announces the software is weak, and its very sign depends on how repositories are paired with
citation counts: undetectable through the sparse naming channel ($n=137$), weakly positive through
repository-declared DOIs ($n=1{,}067$). We report that fragility rather than a single decoupling
estimate; at the linkage density these corpora currently support, the measurement is too
gap-sensitive to carry a strong claim either way. These patterns are tempered by a linkage whose
recall is demonstrably low (\S\ref{sec:validation}), so we read them as motivating, rather than
settling, a software-aware view of scientific impact.
The same instrument lets us follow an emerging mode of production (AI coding agents) into the
S3C; because adoption post-dates the impact stocks, we identify what
\emph{facilitates} it (development intensity and relative youth, not field or popularity) and
defer the long-run outcome study to a future snapshot. The broader
implication is methodological: as the instruments trained on science become operational and
software-mediated, software-aware, dependency-grounded impact must become a first-class
quantity in the science-of-science toolkit, not a term left unmeasured.

\paragraph{On triangulated anchoring.} The representativeness check
(\S\ref{sec:validation}) shows that any single science-software seed is one partial
operationalization; triangulating four independently-constructed anchors (SciCat, JOSS,
Softcite tool names, and a CRAN/conda package registry) confirms they are complementary and
span a grounded-to-connected gradient, and that the reuse--citation coupling is weakly positive
and sign-stable across all of them. The remaining step is to carry the anchor union through the
\emph{full} RQ\,A/RQ\,B leader tables and the field-level analyses (not just the headline
channels reported here), and to replace the CRAN/conda proxy with a curated scientific-package
list. Because the anchors carry different biases (classifier recall, JOSS's post-2016 recency
and its by-construction paper link, registry infrastructure skew), their disagreements localize
where the linkage must improve before absolute magnitudes can be trusted.

\paragraph{Data availability.} The cross-corpus graph (\texttt{edges.typed.gz}, eight typed
layers), the per-lens result tables (\rqA{}/\rqB{} rankings, decoupling and Jaccard statistics,
AI-adoption models), and the build and analysis scripts that reproduce every number reported
here are provided as a replication package, archived at Zenodo (DOI to be assigned). The
underlying World of Code maps (V2604) are available through the World of Code
infrastructure~\citep{woc19}, and the SciCat seed through its
release~\citep{malviyathakur2023scicatcurateddatasetscientific}.

\bibliographystyle{plainnat}
\bibliography{references}

\end{document}